\documentclass[aps,prd,amsmath,amssymb,12pt]{revtex4}
\usepackage{graphicx}% Include figure files
\usepackage{color}
\usepackage{bm}% bold math
\def\journal#1#2#3#4{{#1} {\bf #2}, #3 (#4)}
\newcommand{\be}{\begin{equation}}
\newcommand{\ee}{\end{equation}}
\newcommand{\bea}{\begin{eqnarray}}
\newcommand{\eea}{\end{eqnarray}}
\newcommand{\hf}{\frac12}
\newcommand{\nn}{\nonumber\\}
\def\eq#1{(\ref{#1})}
\def\la{\langle}
\def\ra{\rangle}

\def\Tr{{\mathrm{Tr}}}
\def\ord#1{{\cal O}\left(#1\right)}
\def\mr#1{{\mathrm{#1}}}

\def\hx{\hat x}
\def\hy{\hat y}
\def\hD{{\hat D}}

\def\ih{\frac{i}{\hbar}}

\begin{document}
\title{Instantaneous and dynamical decoherence}
\author{Janos Polonyi}
\email{polonyi@iphc.cnrs.fr}
\affiliation{Strasbourg University, CNRS-IPHC, 23 rue du Loess, BP28 67037 Strasbourg Cedex 2, France}

\begin{abstract}
Two manifestations of decoherence, called instantaneous and dynamical, are investigated. The former reflects the suppression of the interference between the components of the current state while the latter reflects that within the initial state. These types of decoherence are computed in the case of the Brownian motion and the harmonic and anharmonic oscillators within the semiclassical approximation. A remarkable  phenomenon, namely the opposite orientation of the time arrow of the dynamical variables compared to that of the quantum fluctuations generates a double exponential time dependence of the dynamical decoherence in the presence of a harmonic force. For the weakly anharmonic oscillator  the dynamical decoherence is found to depend in a singular way on the amount of the anharmonicity.
\end{abstract}
\maketitle

\section{Introduction}
According to the standard usage of the term decoherence denotes the suppression of interference between certain components of a quantum state \cite{zehd,zurekd}. Thus, decoherence is a fingerprint of an environment since in closed systems the unitary dynamics sustains quantum coherence. Moreover, decoherence is not separable from dissipation, and also a necessary element of the quantum-classical transition \cite{joos,zurekt} and thereby of the recovery of the additive probabilities of histories in the classical limit \cite{gellmann,griffiths,omnes,halliwell}.

In contrast to its numerous significance it became customary to identify decoherence with the suppression of the off-diagonal elements of the reduced density matrix of the observed system during its temporal evolution. The aim of the present work is to point out that the decoherence should be defined in a more careful manner, by paying more attention to the internal system dynamics. We present two alternative signatures, the instantaneous and the dynamical decoherence. The instantaneous decoherence is built on the customary way to identify decoherence and relies on the suppresion of the interference terms within the actual state. The dynamical decoherence confirms the intuitive view about the builing up of the suppression during the time evolution and has not been mentioned before. The particular definition of the instantaneous decoherence was chosen to make the comparison with the dynamical decoherence natural and easy.

The instantaneous decoherence refers to the current state of the observed system by the help of some indicator of the mixed state components, such as the entropy or the purity. In the procedure, followed below, one starts with the specification of two orthogonal pure states, $|\psi_\pm\ra$, $\la\psi_+|\psi_-\ra=0$, and monitors the suppression of the interference terms occurring in the probability of finding the system in the subspace of the pure state, $|\psi\ra=\sum_{\sigma=\pm1}|\psi_\sigma\ra$,
\be
\sum_{\sigma\sigma'}\Tr[|\psi_\sigma\ra\la\psi_{\sigma'}|\rho(t)]\to\sum_{\sigma=\sigma'}\Tr[|\psi_\sigma\ra\la\psi_{\sigma'}|\rho(t)],
\ee
$\rho(t)$ being the current reduced density matrix of the open system considered and the trace is to be taken  with respect to the degrees of freedom of that system. This particular definition is employed to be as close as possible to the dynamical decoherence, defined by the suppression of the interference terms of the initial state in the expectation value of an observable $A$,
\be
\sum_{\sigma\sigma'}\Tr[A(|\psi_\sigma\ra\la\psi_{\sigma'}|)_t]
\to\sum_{\sigma=\sigma'}\Tr[A(|\psi_\sigma\ra\la\psi_{\sigma'}|)_t].
\ee
Here $(|\psi_\sigma\ra\la\psi_{\sigma'}|)_t$ denotes the component $|\psi_\sigma\ra\la\psi_{\sigma'}|$ of the initial density matrix developed until the current time. Both the pure states and the density matrix follow linear time evolution with the important difference that no interference terms appear in observable averages for the latter. The most obvious choice for $A$ is $A=\sum_\sigma|\psi_\sigma\ra\la\psi_\sigma|$. 

To illuminate the conceptual difference between the two type of decoherence introduced above, let us consider Schr\"odinger's cat as an example. Here $|\psi_\pm\ra$ corresponds to the cat being alive or dead and the probability of finding the cat in the living or dead state, $p_\pm$, can be expressed in terms of the initial state, given by the help of $|\psi_\pm\ra$. Dynamical decoherence means the suppression of the interference terms between the two distinct states at the instant of time when the experiment was prepared in the final probability $p_\pm$. Thus, dynamical decoherence displays the loss of informations, encoded in these interference terms during the period of time between preparation and observation. On the other hand, instantaneous decoherence reflects the presence of mixed components in the density matrix at the instant of time of the observation.

The method of investigation to be applied consists of the Closed Time Path (CTP) formalism \cite{schw,keldysh,kamenev,rammer,calzetta}, and the path integral representation for the propagator of the reduced density matrix of an open system. This is a CQCO scheme, i.e. it handles classical, quantum, closed and open systems on equal footing \cite{effth}. The temporal development of the density matrix of a closed system is a unitary transformation, $U\rho_iU^\dagger$, $\rho_i$ being the initial density matrix. The unitary operators $U$ and $U^\dagger$ act in mutually dual bra and ket spaces and generate two independent and equivalent copies of the pure states. The degrees of freedom are redoubled, $x\to(x^+,x^-)$, in the framework of the CTP formalism to represent both copies which become coupled in an open system where one usually aimes at the reduced density matrix, $\Tr_e[U\rho_iU^\dagger]$, the trace being taken with respect to the unobserved degrees of freedom. The observed open system is usually much smaller than its environment and the description of the open system dynamics in terms of interaction between two system copies leads to dramatic simplification. The classical dynamics is recovered by restricting the two copies identical and the quantum fluctuations can be identified as the deviation of the two copies. Transforming the coordinates according to $(x^+,x^-)\to(x,x^d)$, $x=(x^++x^-)/2$, $x^d=x^+-x^-$ proves to be particularly suitable for the task at hand because $\la x\ra$ coincides with the coordinate expectation value and $x^d\to0$ in the classical limit, $\hbar\to0$. The Fourier transform of the density matrix in $x^d$ yields the Wigner function, offering a formal analogy with classical dynamics in the phase space \cite{phasespace} and a description of decoherence, motivated by classical physics \cite{phsspdec}.

Interpreting, in the equation of motion of $x$, $x^d$ as a noise term offers a generalization of the Langevin equation method to quantum systems. For harmonic models, this noise is imaginary, and the analytical continuation of the path integral to imaginary values of $x^d$ leads to real noise and an equivalent representation of quantum transition amplitudes in terms of a Langevin equation \cite{feynman,schmid}. It should be noticed, however, that such a noise is not a fingerprint of an environment. Rather, it occurs also  in closed quantum systems. Quantum Langevin equations for  open systems have been established previously by solving the environment equation of motions in the Heisenberg representation \cite{ford,gardiner}. This procedure is equivalent to applying the CTP formalism, apart from the fact that the latter can handle interactive environment in a much simpler manner. 

Simple toy models \cite{dekker} have already been used to find the impact of a harmonic environment on the observed system \cite{agarwal,calderialeggett,unruh}. The so called stationary decoherence was introduced in \cite{joos} within the framework of kinetic theory, supplemented later by including dephasing and dissipation \cite{diosi,adler}. More systematic investigations used the Born approximation \cite{vacchini,lanz,vacchinie}, taking into account higher orders \cite{hornbergerk}, and utilized the usual many-body methods \cite{dodd}. The prototype of the models used in his work consists of a test particle (the system) interacting with an ideal gas (the environment). The degrees of freedom of the latter are eliminated and the effective Lagrangian is calculated within the leading order of the perturbation expansion with respect to the test particle-gas interaction, and the Landau-Ginzburg double expansion \cite{gas}. The resulting effective Lagrangian is equivalent with the traditional models \cite{calderialeggett}. 

The path integral formalism offers an alternative way to imagine and to deal with quantum systems. The decoherence has been identified and mainly studied in the operator formalism but it is natural to explore the possibilities of using the path integral formalism for its detailed description \cite{grabert}. An important advantage of the path integral formalism, its flexible handling of a non-local effective dynamics, was exploited in the calculation of the non-local, time-dependent form of the master equation for harmonic \cite{humeh} and anharmonic environment \cite{humeah}. Another approach, the consistent history formalism of quantum mechanics \cite{gellmann,griffiths,omnes,halliwell} leads to the decoherence functional \cite{dowker}, a modified form of the influence functional \cite{feynman} of the CTP formalism. The path integral representation is particularly advantageous to find the effects of the coarse graining of the particle trajectory \cite{hartlecg,brun} and to describe continuous monitoring of a quantum system by measurements \cite{menskii}. One can gain a simple insight into the propagation and the decoherence of a relativistic particle \cite{halliwellr} by the help of integrating over the particle trajectory in space-time. The interplay of decoherence and dissipation in front of a dielectric plate, an interesting polarization effect, was addressed in ref. \cite{farias}. The master equation, the traditional description of decoherence, was derived within the harmonic oscillator model in the presence of initial system-environment correlations \cite{romero} and for an electron in QED \cite{anastopoulos}. The decoherence of a particle, subject of a harmonic force and coupled linearly to a harmonic environment, can be followed by solving the local, stationary master equation \cite{lindblad,sandulescu,isar}. The saddle point expansion of the path integral expression for the Liouville-space propagator of the density matrix, introduced in this work, agrees with these results, is a systematical approximation scheme for the decoherence of realistic, anharmonic systems and offers a simple, intuitive picture of the open dynamics.

The semiclassical approximation yields exact solutions for the Brownian motion and the open harmonic oscillator and the $\ord\hbar$ approximation in the anharmonic case. The distinguished property of the harmonic model is the strict separation of the first and the second moments of the canonical variables, the former being controlled by classical physics whereas the latter being shaped by the quantum fluctuations. Owing to the Wick theorem the corresponding sectors in the higher order Green functions remain separate. While dissipation modifies the dynamics of $x$ in a monotonic manner,  decoherence, being expressed by both coordinates, $x$ and $x^d$ reflects inherent non-monotonicity. Another difference between dissipation and decoherence, apparent on the level of the double exponential time dependence is due to the different direction of the dissipative force in time for $x$ and $x^d$. Notice, however, the separation of the dissipation and the decoherence is possible for harmonic models only; anharmonicity couples the first two moments and render so these phenomena inseparable from each other.

An open system exhibits at least two characteristic time regimes; a transient and a relaxed one. The former decouples the initial state from the subsequent development, within the latter the system approaches asymptotically a (quasi-)stationary state. The irreversibility, indicated by the breakdown of the time reversal symmetry of the propagator, is negligible in the transient phase and becomes manifest by the relaxation, providing thereby a clear separation of the instantaneous and the dynamical decoherence. The classical Brownian motion has a single time scale encoded by the friction force. The open harmonic oscillator has three of them and exhibits an intermediate time regime. In this model the dynamical decoherence supports a double exponential time dependence in the relaxed regime. Such a rapid time dependence seems to be ``screened'' by anharmonicity. Thus, the Brownian motion, the harmonic oscillator and the anharmonic oscillator belong, with respect to the decoherence, to three different classes of open dynamics.

The presentation starts with the separation of the instantaneous and the dynamical decoherence in section \ref{ifcompds}, followed by the brief outlines of the semiclassical approximation of the decoherence in section \ref{semicls}. The decoherence of the harmonic toy models is discussed in section \ref{hss} and section \ref{nonhos} contains some remarks about the anharmonic oscillator. A summary is given in section \ref{concls} and a brief justification of the phenomenological Lagrangian, used in the calculation, is given in an appendix.

\section{Signatures of the instantaneous and the dynamical decoherence}\label{ifcompds}
A measure of the instantaneous decoherence of two orthogonal states, $|\psi_\pm\ra$, can easily be identified as the expectation value of the observable $A_o=|\psi_+\ra\la\psi_-|+|\psi_-\ra\la\psi_+|$, $\Tr\rho A_o=2\mr{Re}\la\psi_+|\rho|\psi_-\ra$. This expression has the undesirable feature to depend on the relative phase and the norm of the states $|\psi_\pm\ra$. The dependence on the relative phase can be eliminated by maximizing over the phases, leading to $2|\la\psi_+|\rho|\psi_-\ra|$. The invariance under the change of the norm, $|\psi_\pm\ra\to\lambda_\pm|\psi_\pm\ra$, is reached in two steps. The normalization with the diagonal contributions, $\Tr\rho A_d$, with $A_d=|\psi_+\ra\la\psi_+|+|\psi_-\ra\la\psi_-|$,
\be
\frac{2|\la\psi_+|\rho|\psi_-\ra|}{\Tr\rho A_d}=\frac{2|\la\psi_+|\rho|\psi_-\ra|}{\la\psi_+|\rho|\psi_+\ra+\la\psi_-|\rho|\psi_-\ra},
\ee
establishes the independence under the common rescaling $\lambda_+=\lambda_-$. The dependence on different rescaling can be eliminated by replacing the arithmetic mean of the diagonal contributions by their geometrical mean,
\be\label{cc}
D_{inst}=\frac{|\la\psi_+|\rho|\psi_-\ra|}{\sqrt{\la\psi_+|\rho|\psi_+\ra\la\psi_-|\rho|\psi_-\ra}}.
\ee
This ratio satisfies the inequality, $0\le D_{inst}\le1$, the pure states saturating the upper bound.

To find a measure of the dynamical decoherence we assume that the system joined with its environment forms a closed full system with Hamiltonian $H_{tot}$ in a factorisable initial state $\rho_{tot}(t_i)=\rho(t_i)\otimes\rho_{ei}$, and write the reduced density matrix at time $t$ as a linear expression of the initial value,
\be\label{rhoschr}
\rho(\hx,t)=\int d\hx_i{\cal G}(\hx,\hx_i,t-t_i)\rho(\hx_i,t_i),
\ee
where $\hx=(x^+,x^-)$ stands for a pair of system coordinates,
\be\label{liouvprop}
{\cal G}(\hx,\hx_i,t-t_i)=\Tr_e[\la x^+|U(t-t_i)| x_i^+\ra\rho_{ei}\la x_i^-|U^\dagger(t-t_i)|x^-\ra],
\ee
denotes the Green-function in the Liouville space, $U(t)=\exp-itH_{tot}/\hbar$, and $\Tr_e$, the trace over the environment. The interference terms of the initial state, given by the component $\Delta\rho(t_i)$ of $\rho(t_i)$, develops into
\be
\int d\hx_id\hx_f|x_f^-\ra\la x_f^+|{\cal G}(\hx,\hx_i,t-t_i)\la x^+_i|\Delta\rho(t_i)|x_i^-\ra.
\ee

The characterization of the weight of this component within the actual state, is a non-trivial  task owing to the unitarity of the full dynamics which suppresses its contribution to the total probability. In fact, this contribution, $\Tr[U(t-t_i)A_o\otimes\rho_{ei}U^\dagger(t-t_i)]$, is independent of $t$ and is vanishing for $t=t_i$. In other words, the interference contributions of the initial state are completely dispersed within the full system as far as the total probability is concerned and can only be recovered by measuring an appropriately chosen observable. It is natural to choose an initial pure state, $\rho(t_i)=|\psi\ra\la\psi|$, $|\psi\ra=|\psi_+\ra+|\psi_-\ra$ and inquire about the probability for finding the state within the subspace span by the components $|\psi_\pm\ra$, $\Tr[A_dU(t-t_i)(A_o+A_d)\otimes\rho_{ei}U^\dagger(t-t_i)]$, up to the normalization. This quantity is the sum of the diagonal and the off-diagonal components of the initial state whose ratio,
\be
\frac{\Tr[A_dU(t-t_i)A_o\otimes\rho_{ei}U^\dagger(t-t_i)]}{\Tr[A_dU(t-t_i)A_d\otimes\rho_{ei}U^\dagger(t-t_i)]},
\ee
is a measure of the suppression of the interference terms of the initial state during the time evolution from $t_i$ to $t$. The replacement of the arithmetic means by geometrical one produces the dynamical suppression factor,
\be\label{dynds}
D_{dyn}=\frac{\prod_{\sigma_f\sigma_i=\pm}\Tr_e[\la\psi_{\sigma_f}|U|\psi_{\sigma_i}\ra\rho_{ei}\la\psi_{-\sigma_i}|U^\dagger|\psi_{\sigma_f}\ra]}{\prod_{\sigma_i\sigma_f=\pm}\Tr_e[\la\psi_{\sigma_f}|U|\psi_{\sigma_i}\ra\rho_{ei}\la\psi_{\sigma_i}|U^\dagger|\psi_{\sigma_f}\ra]}.
\ee
It detects the correlation between the operators $U$ and $U^\dagger$ in the expectation values, the presence of mixed components of the state of the system at time $t$. For closed system $D_{dyn}=1$.

\section{Semiclassical decoherence}\label{semicls}
The transition amplitude of a closed system,
\be
\la x_f|U(t)|x_i\ra=\int D[x]e^{\ih S[x]},
\ee
is found by integrating a phase factor over the trajectories with end points $x(t_i)=x_i$, $x(t_f)=x_f$. One can similarly write the Liouville-space propagator of the density matrix, $\la x^+|U(t-t_i)| x_i^+\ra\la x_i^-|U^\dagger(t-t_i)|x^-\ra$, as a path integral, 
\be\label{lpropeffa}
{\cal G}(\hx_f,\hx_i,t)=\int_{\hx(t_i)=\hx_i,\hx(t_f)=\hx_f}D[\hx]e^{\ih(S[x^+]-S[x^-])}.
\ee
over a pair of open trajectories pairs, $\hx=(x^+,x^-)$, with end points $\hx(t_1)=\hx_i$, and $\hx(t_2)=\hx_f$. The specification of all degrees of freedom along the trajectories leaves no room for diffraction and the contributions are phase factors with unit modulus. In case of an open system we write the total action as $S[x,y]=S_s[x]+S_e[x,y]$ where $y$ denotes the environment coordinate and assume that there is no system-environment entanglement in the initial state at $t_i$. The integration over the environment trajectories yields the expression
\be\label{lsprop}
{\cal G}(\hx_f,\hx_i,t)=\int_{\hx(t_i)=\hx_i,\hx(t_f)=\hx_f}D[\hx]e^{\ih S_{eff}[\hx]},
\ee
for the propagator \eq{liouvprop}, including the effective action, 
\be
S_{eff}[\hx]=S_s[x^+]-S_s[x^-]+S_{infl}[\hx],
\ee
defined by the help of the influence functional \cite{feynman},
\be
e^{\ih S_{infl}[\hx]}=\int_{y^+(t_f)=y^-(t_f)}D[\hy]e^{\ih S_e[\hx,\hy]},
\ee
where the integration is taken over closed paths, $y^+(t_f)=y^-(t_f)$, to incorporate the trace operation in the definition of the reduced density matrix. Note that the integration in the Liouville-space propagator is over the closed paths of the environment and the open paths of the observed system. We do not have access to all dynamical degrees of freedom in the case of an open system and the diffraction processes, taking place within the unobserved environment, can suppress the magnitude of the contribution of a given (system) trajectory to the (reduced) density matrix and $\mr{Im}S_{eff}[\hx]\ne0$. In other words, the decoherence is encoded by $\mr{Im}S_{eff}$, the suppression of the contribution of a pair of trajectories.

Note that the full time reversal transformation, $x^\pm\to x^\mp$ and $S_{eff}[x^-,x^+]=-S^*_{eff}[x^+,x^-]$, exchanges the direction of the time together with the initial and the final conditions hence is always a trivial, formal symmetry. Another important feature of the effective action expresses the unitarity of the full dynamics, $\Tr\rho=1$. This condition becomes highly non-trivial by introducing a physical external source, coupled to an observable, $H_{tot}\to H_{tot}+j(t)A(t)$, and considering $Tr[\rho]$ as the generator functional for the Green functions for $A$. In particular, when the system moves along diagonal CTP trajectories, $x^+(t)=x^-(t)$, then it represents a given, possible classical environment for its environment and the unitarity of its dynamics, $\Tr\rho_e=1$ where $\rho_e$ is the environment density matrix, implies $S_{eff}[x,x]=0$. 

The Liouville space propagator, \eq{lsprop}, is approximated below in an illuminating manner by a combination of phenomenological considerations and the expansion in powers of the Planck constant. The former is used to define a simple, physically motivated influence functional and the latter consists of the semiclassical approximation when truncated at $\ord\hbar$.

\subsection{Phenomenological effective Lagrangian}
The usual way to find a local effective action is the Ginzburg-Landau local expansion with the assumption of the smallness of the amplitude and the frequency of the modification of the quantum trajectories by the environment. To construct the leading order, harmonic Lagrangian, we possess 10 possible terms, the bilinears made by $x$, $\dot x$, $x^d$, and $\dot x^d$, whose coefficients are real or pure imaginary numbers owing to the full time reversal invariance. The vanishing of the action for $x^d(t)=0$ eliminates the combinations $x^2$, $\dot xx$ and $\dot x\dot x$, allowing $xx^d,\dot xx^d,x\dot x^d,\dot x\dot x^d$ with real coefficients and imaginary numbers, multiplying $x^{d2}$, $\dot x^dx^d$ and $\dot x^{d2}$. The total time derivatives, $\dot x^dx^d$, $\dot xx^d+x\dot x^d$, drop out from the equations of motion playing however a role in quantum mechanics. The term $\dot xx^d+x\dot x^d$ generates a gauge transformation, a basis transformation, and $\dot x^dx^d$ changes the decoherence strength. Both influence the effective action in a trivial manner and will be ignored. One may add arbitrary local potentials without creating much trouble in the initial phase of formal calculations. Therefore we start with the effective Lagrangian, 
\bea\label{efflagr}
L_{eff}&=&\frac{m}2(\dot x^{+2}-\dot x^{-2})+\frac{k}2(\dot x^+x^--\dot x^-x^+)-U(x^+)+U(x^-)\nn
&&+i\left[V(x^+-x^-)+\frac{d_2}2(\dot x^+-\dot x^-)^2\right],
\eea
cf. Appendix \ref{effls} for more justification. The Lagrangian assumes the form
\be
L_{eff}=m\dot x\dot x^d-\frac{k}2(\dot xx^d-x\dot x^d)-U\left(x+\frac{x^d}2\right)+U\left(x-\frac{x^d}2\right)+i\left[V(x^d)+\frac{d_2}2\dot x^{d2}\right]
\ee
in the parametrization $x^\pm=x\pm x^d/2$.

The imaginary part of the Lagrangian merits a special attention in discussing decoherence. The effective action of an open system can be defined in classical mechanics, as well \cite{effth}. Since one is interested in CTP diagonal trajectories in classical physics, $x^+=x^-$, the equation of motion for $x$ imposes $x^d=0$. Such a restriction on the environment coordinates suppresses the imaginary part of the influence functional, leaving an infinitesimal, $\ord{\epsilon}$, imaginary part of the effective action which incorporates the $\epsilon$-prescription for the retarded and advanced Green functions of the system. However, the irreversibility of the open dynamics appears through negative time parity terms in the influence functional, c.f. the second, friction term of the Lagrangian \eq{efflagr} \cite{bateman}. The imaginary part of the effective action of a closed quantum system remains $\ord\epsilon$ like in the usual path integral representation of the transition amplitude between pure states. The construction of the effective dynamics in the presence of the environment can be considered as a coarse-graining and the information loss generates an imaginary part to the influence functional and entropy. Furthermore,  the interference between different environment states generates $\ord{\epsilon^0}$ imaginary part to the effective action. The harmonic part of $\mr{Im}S_{eff}$ and the dissipative terms of $\mr{Re}S_{eff}$ replace the formal $\epsilon$-prescription by shifting the poles of the Green functions off the real frequency axes and giving rise of finite life-times and decoherence. Note that dissipation and decoherence are already present in an infinitesimal extent within the closed dynamics under the disguise of the $\epsilon$-prescription and their finite presence in open systems can formally be regarded as a spontaneous symmetry breaking \cite{irr}.

\subsection{Stationary decoherence}
A very simple approximation of the path integral \eq{lsprop} is the replacement by its integrand taken at some physically motivated trajectory, $x^\pm(t)$. The estimate of the decoherence by the help of $\mr{Im}S_{infl}[\hx]$, evaluated along the chosen trajectory, can be called rigid decoherence because the system dynamics is completely ignored. In the simplest rigid scheme the pair of trajectories is taken taken to be stationary, $x^\pm(t)=x^\pm_0$, leading to stationary decoherence \cite{joos}. Note that strong decoherence, displayed by systems with weak internal interactions, compared with the system-environment interactions, can be approximated by the rigid decoherence only if the dominance of the path integral \eq{lsprop} by the considered trajectory pair is established and the stationary trajectories may loose their importance even in weakly interactive systems. The path integral, \eq{lsprop}, approximated by the integrand at the trajectory $x^\pm(t)=x\pm x^d/2$ yields
\be
{\cal G}_{st}(\hx_f,\hx_,t)=e^{-i\frac{t}{2\hbar}[U(x+\frac{x^d}2)-U(x-\frac{x^d}2)]-\frac{t}\hbar V(x^d)},
\ee
where $t=t_f-t_i$. The resulting stationary decoherence time scale, $\tau_{sd}(x^d)=\hbar/V(x^d)$, depends on $x^d$. One can always find a characteristic stationary decoherence length scale, $\ell_{sd}$, by dimensional reasoning, in particular the harmonic decoherence potential, $V(x^d)=d_0x^{d2}/2$, yields $\ell^2_{sd}=2\hbar/d_0t$. Note that $\tau_{sd}(x^d)$ is not physical since $x^d$ being non-observable, $\la x^{dn}\ra=0$. Indeed, consider the trace of the density matrix in the presence of a linear source, $j(t)$, coupled to the coordinate $x(t)$,
\be
Z[j]=\Tr[T[e^{-\ih\int^{t_f}_{t_i}[H-j(t)x(t)]}\rho(t_i)T^*[e^{\ih\int^{t_f}_{t_i}[H+j(t)x(t)]}],
\ee
where $T$ and $T^*$ denotes the time and the anti-time ordering. In the path integral formula the source is coupled to $x^d=x^+-x^-$,
\be
Z[j]=\int D[\hx]e^{\ih S_{eff}[\hx]-\ih\int dtj(t)x^d(t)},
\ee
and the moments, $\la x^{dn}(t)\ra=(-i\hbar)^n\delta^mZ[0]/\delta j(t)^n$, are vanishing because the unitarity of the time evolution imposes $Z[j]=0$. Nevertheless the stationary time scale may be useful since its minimal value, $\min_{x^d}\tau_{sd}(x^d)$, represents a lower bound on other decoherence time scales calculated in the semiclassical approximation.

\subsection{Semiclassical approximation}
The saddle point expansion of the path integral \eq{lsprop} represents a systematic approximation scheme. The leading order contribution is given by the integrand, evaluated at the trajectory which solves the Euler-Lagrange equation of motion. The next order produces a multiplicative factor, representing the fluctuations,
\be\label{propdmho}
{\cal G}(\hx_f,\hx_,t)={\cal N}(\hx_f,\hx_,t)e^{\ih S_{eff}(\hx_f,\hx_i,t)}.
\ee
In the case of harmonic system this equation is exact and the normalization, $\cal N$, depends on the time only. The saddle point trajectory satisfies the equations,
\bea\label{eomxxd}
m\ddot x&=&-\hf U'\left(x+\frac{x^d}2\right)+\hf U'\left(x-\frac{x^d}2\right)-k\dot x+i[V'(x^d)-d_2\ddot x^d]\nn
m\ddot x^d&=&-U'\left(x+\frac{x^d}2\right)+U'\left(x-\frac{x^d}2\right)+k\dot x^d,
\eea
together with the boundary conditions, $\hx(t_i)=\hx_i$ and $\hx(t_f)=\hx_f$. The following remarks are in order at this point: (i) The saddle point trajectory is made complex by the decoherence and the quantum fluctuations, $x^d$, act as a complex noise on the physical coordinate, $x$. This highlights an additional role of decoherence: the quantum fluctuations appear as a noise in the dynamics via the decoherence. The saddle point trajectory $x^d(t)$ is real for a harmonic potential, $U(x)=\ord{x^2}$, and the noise for the physical coordinate is imaginary. It is easy to continue analytically the integration over $x^d(t)$ to arrive at a Langevin equation with real noise \cite{feynman,schmid}, providing thereby an equivalent derivation of the path integral results for harmonic models. The quantum Langevin equation can be derived in the Heisenberg representation of the operator formalism \cite{ford,gardiner}. This gives yet another equivalent treatment of the dynamics of open harmonic models. The anharmonic terms in the potential $U(x)$, treated in the leading order saddle point approximation, renders the saddle point trajectory $x^d(t)$ complex similar to the noise in the Langevin equation for the physical coordinate. (ii) There are non-trivial stationary solutions,
\be\label{statp}
iV'(x^d)=U'\left(x+\frac{x^d}2\right)=U'\left(x-\frac{x^d}2\right)
\ee
balancing the complexified Newtonian force with the noise of the Langevin scheme and they may be important in forming the relaxed asymptotic state. (iii) The ``wrong'' sign of the friction force in the equation of motion for $x^d$ makes $x^d(t)$ a runaway trajectory which can be stable by the final condition, $x^d(t_f)=x_f^d$, only. This instability disappears in the limit $\hbar\to0$, and $x^d(t)=0$ is recovered in the classical CTP formalism \cite{arrow,galley}. Thus the quantum fluctuations are unstable; they have the opposite time arrow as compared to the physical variables. This feature destabilizes some of the stationary points, the solutions of eqs. \eq{statp}.

\section{Harmonic systems}\label{hss}
In the case of a harmonic system, $U(x)=m\omega^2x^2/2$, $V(x^d)=d_0x^{d2}/2$ the real part of the Lagrangian has three classical parameters, the mass, the oscillator frequency and the friction constant which determine the trajectory in the classical, $x^d\to0$, case. The imaginary part contains two parameters, describing velocity independent and velocity dependent decoherence, $d_0$ and $d_2$, respectively. The coordinates $\mr{Re}x$ and $x^d$ satisfy the equation of motion of a classical, damped oscillator with oppositely running time whose solution contains the normal frequencies $\omega_{ss'}=si\nu_0/2+s'\omega_\nu$, where $\omega_\nu=\sqrt{\omega^2-\nu^2/4}$, $\nu=k/m$ with $(s,s')=(+,\pm)$ and $(s,s')=(-,\pm)$, respectively. $\mr{Im}x$ satisfies a similar equation of motion except that it is driven by $x^d$. The solution of such an equation contains all the four normal frequencies. It is remarkable that the time dependence of the first moments of the canonical variables, $x$ and $p$, is described by the normal frequencies, known from the classical oscillator. Hence the characteristic times are given by classical physics, and the decoherence parameters of the Lagrangian appear as multiplicative constants the time dimension of which is removed in the saddle point trajectories by the classical parameters $\omega$ or $\nu$ rather than the time itself.  

The saddle point trajectory is of the form $x^\sigma(t)=\sum_{\sigma'ss'}[c_{ss'i}^{\sigma\sigma'}e^{i\omega_{ss'}t}x^{\sigma'}_i+c_{ss'f}^{\sigma\sigma'}e^{i\omega_{ss'}t}x^{\sigma'}_f]$, and the coefficients are rational polynomials of the exponential factors $\exp i(t_i-t_f)\omega_{s,s'}$. It displays a transient $t\ll\tau_i$, an intermediate $\tau_i\ll t\ll\tau_r$ and a relaxed $\tau_r\ll t$ time regime, where $1/\tau_i=\max_\sigma(\mr{Im}\omega_{+\sigma})$ and $1/\tau_r=\min_\sigma(\mr{Im}\omega_{+\sigma})$. The action is a quadratic expression of the initial and final coordinates with coefficients, given by rational polynomials of $\exp i(t_i-t_f)\omega_{s,s'}$. The intermediate time regime shrinks to zero in the case of the Brownian motion, $\omega_0\to0$, leaving two non-vanishing normal frequencies, $\pm\nu$, and a double degenerate vanishing frequency. This latter generates a polynomial dependence in $t$ and $t_f-t_i$ in the coefficients. 

The effective action, evaluated for the saddle point trajectories, is quadratic in the initial and final points and can be written in the generic form
\bea\label{effactho}
S_{eff}(\hx_f,\hx_i,t)&=&\frac{M}t(x_f-x_i)(x_f^d-x_i^d)-t\frac{M\Omega^2}4(x_f+x_i)(x^d_f+x^d_i)-\frac{K}2(x_f^d+x_i^d)(x_f-x_i)\nn
&&+i\left(\frac{D_i}2x^{d2}_i+\frac{D_f}2x^{d2}_f+D_mx^d_ix_f^d\right),
\eea
in terms of time dependent parameters. The normalization, $\cal N$, is fixed by $\Tr\rho=1$. The effective action with $D_i=D_f$ is the trivial generalization of the Lagrangian, \eq{efflagr}, obtained by replacing the time derivatives by finite differences. However, the possibility $D_i\ne D_f$ is needed to take into account the renormalization of the imaginary boundary term. A trivial but lengthy calculation of the saddle point action can be summarized by listing the expressions
\bea\label{cpmcif}
M&=&m\frac{\omega_\nu t}{\sin\omega_\nu t}\frac{\cos\omega_\nu t+\cosh\frac{\nu t}2}2,\nn
\Omega^2&=&\frac{4(\cosh\frac{\nu t}2-\cos\omega_\nu t)}{t^2(\cosh\frac{\nu t}2+\cos\omega_\nu t)},\nn
K&=&2m\omega_\nu\frac{\sinh\frac{\nu t}2}{\sin\omega_\nu t},
\eea
and
\bea\label{lpmlif}
D_{\stackrel{i}{f}}&=&\frac{\pm\tilde d_+[4(\omega_\nu^2e^{\pm\nu t}-\omega^2)+\nu^2\cos2\omega_\nu t]-2\tilde d_-\omega_\nu\nu\sin2\omega_\nu t}{8\omega^2\nu\sin^2\omega_\nu t},\nn
D_m&=&\frac{\omega_\nu(\tilde d_-\nu\sin\omega_\nu t\cosh\frac{\nu t}2-2\tilde d_+\omega_\nu\cos\omega_\nu t\sinh\frac{\nu t}2)}{2\omega^2\nu\sin^2\omega_\nu t},
\eea
with $\tilde d_\pm=d_0\pm d_2\omega^2$. The real part of the effective action, given by the parameters \eq{cpmcif}, is classical and of $\ord{\hbar^0}$, tree-level, and the imaginary part, containing \eq{lpmlif}, is the effect of quantum fluctuations since $d_0$ and $d_2$ contain the impact of the quantum fluctuations in the environment. The divergences at (half) integer periods are the remnant of the (anti) focusing of the undamped oscillator. 

The dynamical suppression factor, \eq{dynds}, corresponding to localized states, $\psi_\pm(s)=\delta(x-x^\pm)$,
\be\label{dynsf}
D_{dyn}=e^{-\frac{D_i}{2\hbar}(x^+-x^-)^2},
\ee
defines the dynamical decoherence length $\ell_{dd}=\sqrt{2\hbar/D_i}$ which is infinite for closed dynamics, any non-triviality being due to the openness of the system. The instantaneous decoherence factor is defined by the reduced density matrix,
\be
\rho(x^+_f,x^-_f;t_f)=\int dx^+_idx^-_i{\cal G}(x_f^+,x_f^-,x_i^+,x_i^-,t_f-t_i)\rho(x^+_i,x^-_i;t_i).
\ee
The perfectly localized state of the continuous spectrum is non-physical since it can not develop diffraction. This shortcoming will be avoided below by considering a wave packet with finite width,
\be\label{rho}
\rho(x,x^d,t)=Ne^{-\frac{q^2(t)}2x^2-\frac{r^2(t)}2x^{d2}+is^2(t)xx^d},
\ee
with $q\le2r$ for the initial state. The parameters $q$, $r$ and $s$ of the actual state can be expressed in terms of the initial values and the parameters \eq{cpmcif}-\eq{lpmlif}. The instantaneous suppression factor, \eq{cc}, is easy to find,
\be\label{instd}
D_{inst}=e^{-\frac{(x^+-x^-)^2}{2\ell_{id}^2}},
\ee
with the asymptotic instantaneous decoherence length, given by $1/\ell_{id}^2=q^2(\kappa^2-1/4)$. The ratio, $\kappa=r/q\ge1/2$, can be considered as a measure of the mixed components of the state. $\kappa=1/2$ corresponds to the pure wave packet.

\subsection{Brownian motion}\label{browns}
Consider first the translation invariant Brownian motion, $\omega=0$, when $\mr{Re}x$ and the real $x^d$ describe two free motions, subject of the same friction force but having opposite time arrows. Hence the single time scale, $\tau_{Br}=\tau_i=\tau_r=1/\nu$, characterizes both the relaxation (dissipation) and the runaway (decoherence) time dependence. $\mr{Re}x(t)$ reaches the vicinity of its final point within a time $\tau_{Br}$. The off-diagonality, described by $x^d(t)$, follows the same behavior backward in time and $x^d(t)\sim x_i^d$ apart of the last $\tau_{Br}$ time interval. Thus the decoherence suppression factor of the stationary decoherence  scenario is recovered with a relative error $\tau_{Br}/t$. The imaginary part, $\mr{Im}x(t)$, is driven by the decoherence potential, $V(x^d)$, it has vanishing initial and final values, hence it is determined by the boundary conditions for $x^d$. It reaches the velocity $d_0x^d/k$ after the time $\tau_{Br}$, assuming $t\gg\tau_{Br}$, and returns to zero in the second part of the motion. 

A simple calculation yields
\bea\label{bmpar}
D_i&=&d_0\frac{2t\nu-3+4e^{-t\nu}-e^{-2t\nu}}{2\nu(1-e^{-t\nu})^2}+d_2\nu\frac{1-e^{-2t\nu}}{2(1-e^{-t\nu})^2},\nn
D_f&=&d_0\frac{1-4e^{-t\nu}+e^{-2t\nu}(2t\nu+3)}{2\nu(1-e^{-t\nu})^2}
+d_2\nu\frac{1-e^{-2t\nu}}{2(1-e^{-t\nu})^2},\nn
D_m&=&d_0\frac{1-2t\nu e^{-t\nu}+e^{-2t\nu}}{2\nu(1-e^{-t\nu})^2}
-d_2\nu\frac{1-e^{-2t\nu}}{2(1-e^{-t\nu})^2},
\eea
where the linear and the exponential time dependence generate two time regimes, a transient and a relaxed phase, separated by $\tau_{Br}$. The approximate equation, $D_i\sim D_f$, valid for $t\ll\tau_{Br}$, reflects the approximate time reversal invariance in the transient regime. The expansion of the exponential functions in this regime yields an $\ord{t^{-1}}$ time dependence, the rapid drop being due to the artifact of a perfectly localized initial state. Different instantaneous measures of the mixed components may differ in their quantitative time dependence, for instance a Zeno-like effects cancels the time dependence of the purity at short time \cite{kim}. The parameters develop a different approximate form in the relaxed phase where the linear time factor of the numerator generates an $\ord{t}$ dependence for $D_i$. That time dependence remains suppressed in $D_f$ and $D_m$ which follow the $\ord{t^0}$ asymptotic. Thus $D_f$ strongly deviates form $D_i$ in this regime, signaling the onset of irreversibility.

\subsection{Oscillator}\label{hos}
The restoring force towards the equilibrium position generates new features. The simultaneous presence of the restoring and the friction force requires overshooting to have saddle point trajectories with given end points. This is significant because overshooting generates a time-dependence of a qualitatively new functional form in the intermediate time regime and renders the stationary decoherence approximation invalid. In particular, some of the parameters \eq{cpmcif}-\eq{lpmlif} grow exponentially in time, the second exponential in the saddle point suppression factor. As well known that the correlation function of the classical diffusion process contains a double exponential time dependence, $\exp(-c'\exp(-t/\tau_{diss}))$ with $c'>0$ \cite{kubo}. The double exponential of the decoherence suppression is $\exp(-x^{d2}\exp(t/\tau_{dd\infty})/\ell^2_{dd\infty})$, where the change of sign in the second exponential is in agreement with the previous observation that the quantum fluctuations and the physical coordinates sens time flow in opposite direction, and defines thereby the decoherence time scale. The asymptotic time dependence of the effective parameters is a power law for short time and the long time dependence is  $\Omega\sim\ord{t^{-1}}$, $M,K,\sqrt{D_i},D_f,D_m\sim\exp\nu t/2$ and $\exp(\nu/2-\bar\omega)t$ with $\bar\omega=\sqrt{\nu^2/4-\omega^2}$, for underdamped and overdamped oscillator, respectively, apart of the oscillations in the former case. 

\begin{figure}
\includegraphics[scale=0.4]{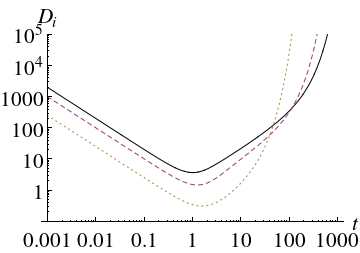}\hskip.1cm\includegraphics[scale=0.4]{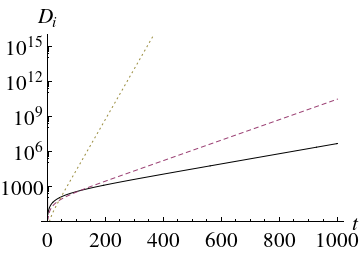}\hskip.1cm\includegraphics[scale=0.4]{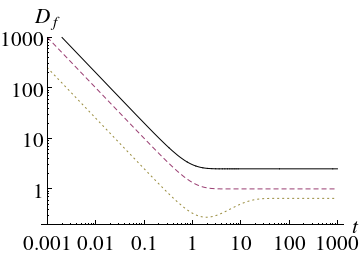}
\centerline{(a)\hskip5cm(b)\hskip5cm(c)}
\caption{The decoherence parameters $D_i$ and $D_f$, plotted against the time with $\omega=0.1$, $\nu=d_0=d_2=2$ (solid line), $\nu=d_0=d_2=1$ (dashed line) and $\nu=d_0=d_2=0.25$ (dotted line), (a): $D_i$ on log-log plot, (b) $D_i$ on log plot and (c) $D_f$ on log-log plot.}\label{dpmf}
\end{figure}

It is instructive to follow the time dependence of the parameters $D_i$ and $D_f$ for different system-environment coupling strength, $g$, i.e. $\nu\to g^2\nu$, $d_0\to g^2d_0$ and $d_2\to g^2d_2$, shown in Fig. \ref{dpmf} in units $\hbar=m=1$. Let us start with the transient regime of Fig. \ref{dpmf} (a) and (c) where the effective parameters of the Brownian motion and the harmonic oscillator are similar and follow power laws. The effective decoherence parameters decrease with increasing coupling strength, $g$. The decrease of $D_i$ as the function of the time indicates the washing out of the quantum information contained in the initial state. The time dependence in the intermediate regime is similar as in the relaxed regime of the Brownian motion: a power law in time with increasing dynamical decoherence and the stationary decoherence picture is valid. In the final, relaxed phase the effective parameters show an exponential increase with slope $\nu$ and $\nu-2\bar\omega$ for underdamped and overdamped oscillator, respectively, c.f. Fig. \ref{dpmf} (b), with no analogy seen in the Brownian motion. The comparison of $D_i$ and $D_f$, shown in Fig. \ref{dpmf} (a) and (c), indicates that the approximative time reversal invariance of the transient regime is strongly violated in the intermediate and the relaxed phases. The dynamical decoherence length is $\ell_{dd}(t)=\ell_{dd}(0)(1+t/\tau_{dd0})+\ord{t^2}$ in the transient regime where the non-universal time scale $\tau_{dd0}$ contains the parameters of the initial state and the Lagrangian. The  asymptotic long time dependence in the relaxed phase is double exponential, mentioned above, with the universal time scale, $\tau_{dd\infty}=1/\nu$ and $1/(\nu-2\bar\omega)$ for underdamped and overdamped oscillator, respectively.

\begin{figure}
\includegraphics[scale=0.4]{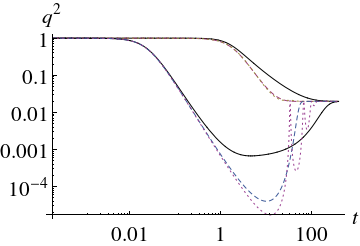}\hskip.1cm\includegraphics[scale=0.4]{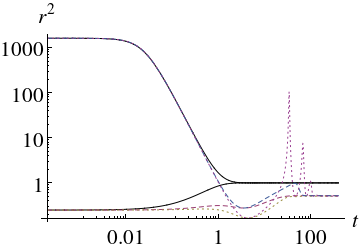}\hskip.1cm\includegraphics[scale=0.4]{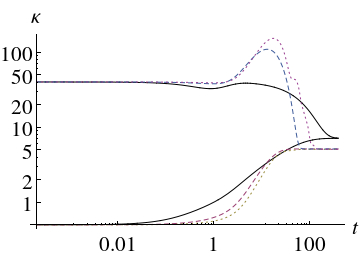}
\centerline{(a)\hskip5cm(b)\hskip5cm(c)}
\caption{The parameters of the wave packet, (a): $q^2$, (b): $r^2$, (c) $\kappa=r/q$, plotted against the time for $\omega=0.1$, $q(t_i)=1$, $\nu=d_0=d_2=1$ (solid line), $\nu=d_0=d_2=0.08$ (dashed line) and $\nu,=d_0=d_2=0.01$ (dotted line). Each function is shown for a pure initial state, $\kappa(t_i)=1/2$, and a mixed wave packet, $\kappa(t_i)=40$. $q^2$ and $r^2$ are decreasing and increasing with $\kappa(t_i)$ in the shown cases, respectively.}\label{qref}
\end{figure}

The instantaneous decoherence is extracted from the parameters $q^2$ and $r^2$ of a wave packet and their combination, $\kappa=r/q$, whose time dependence is shown in Fig. \ref{qref}. One can easily recognize the transient regime where a closer look revels a weak power law time dependence, followed by an intermediate regime where the shift towards the asymptotic values starts and terminates with the relaxed regime with exponentially fast convergence. The parameters approach their asymptotic, relaxed values, $q^2_\infty=2m^2\nu\omega^2/\hbar\tilde d_+$, $r^2_\infty=(\tilde d_+^2+d_0d_2\nu^2)/2\hbar\nu\tilde d_+$ which yield the asymptotic instantaneous decoherence length,
\be
\ell_{id\infty}=\sqrt{\frac{2\hbar\nu\tilde d_+}{\tilde d_+^2+\nu^2(d_0d_2-m^2\omega^2)}},
\ee
a real number if the inequality, $\nu^2\le2d_0d_2/m^2$, needed to assure the positivity of the density matrix \cite{gas}, is satisfied. The decoherence length approaches its asymptotic value with the same time scale as in the case of the dynamical decoherence. The short time dependence in the transient regime is linear and non-universal for our particular definition of the instantaneous decoherence, $\ell_{id}(t)=\ell_{id}(0)(1+t/\tau_{id0})+\ord{t^2}$, as for the dynamical decoherence.

\section{Anharmonic oscillator}\label{nonhos}
The opposite time arrow of the physical coordinate and its quantum fluctuations, together with the imaginary parts of the saddle point trajectory lead to a characteristic difference between the dynamics of the harmonic and the anharmonic open systems. The impact of the opposite time arrows is easiest to see in the relaxed, asymptotic state. For this purpose let us place an anharmonic system into an initial state which is localized around a stable equilibrium position in such a manner that the harmonic approximation is justified for a short time span. The saddle point of the quantum fluctuation, $x^d(t)$, being unstable, drives the system away from the initial region of harmonicity. Such a runaway motion can be stabilized by anharmonicity. It is well known that the saddle point, being the result of an equilibrium between  harmonic and anharmonic forces, $m\omega^2x=gx^n$ with $n>1$, is singular in the limit where the coupling strength, $g$, approaches zero. Hence the relaxed state is non perturbative, the limit of vanishing anharmonicity is not continuous. Another manifestation of this phenomenon is that the saddle point trajectory wanders around the unstable fixed points, \eq{statp}, in a rather complicated manner, controlled by the boundary conditions.

One can gain more insight into the build up of instability by the anharmonicity by recalling the conjecture that the perturbation expansion is singular in quantum systems. This feature of the perturbation expansion has been put forward first in QED \cite{dyson}. The heuristic argument for anharmonic oscillator, $U(x)=m_0\omega_0^2x^2/2+gx^4/4!$, starts by extending $g$ to complex values. Were the radius of convergence, $r_c$, finite then the perturbation expansion would converge for $|g|<r_c$. But this is not possible because there is no ground state if $g<0$. The small amplitude classical motion remains regular for $g<0$ and the dangerous secular contributions can be dealt with in the perturbation expansion. However, this is not enough in quantum mechanics where tunneling always opens up an instability. Another view of such an instability is offered by an open anharmonic quantum system treated in the semiclassical approximation, where the decoherence complexifies the saddle point and makes the singularity to appear already at $g=0^+$. In fact, a term $x^n$ of the potential $U(x)$ generates a time reversal invariant Newtonian force $(-1)^{n/2}n(\mr{Im}x)^{n-1}$ or $(-1)^{n/2}2^{1-n}n(\mr{Im}x^d)^{n-1}$ in the equation of motion for $\mr{Im}x$ or $\mr{Im}x^d$, cf. eqs. \eq{propdmho}, leading to unstable, runaway trajectories for even $n/2$.  Such an instability is known in closed systems where Feynman's $\epsilon$-prescription, $d_0=\epsilon$, can be regarded as the effect of a weak decoherence, to be removed after solving the dynamical problem.

\begin{figure}
\includegraphics[scale=0.4]{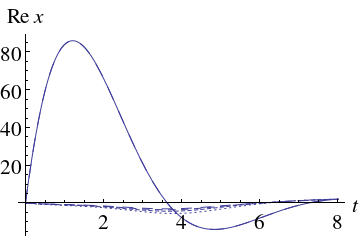}\hskip1cm\includegraphics[scale=0.4]{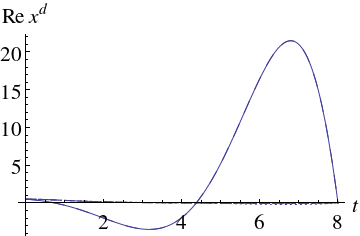}
\centerline{(a)\hskip5cm(b)}
\includegraphics[scale=0.4]{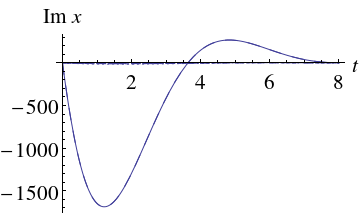}\hskip1cm\includegraphics[scale=0.4]{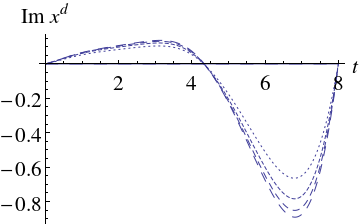}
\centerline{(c)\hskip5cm(d)}
\caption{The complex saddle point, plotted against the time for an underdamped anharmonic oscillator, $\omega_0=1$, $x_i=0$, $x_i^d=0.5$, $x_f=2$, $x_f^d=0$. The coupling constant changes from $g=0$ till $g=0.01$ in an equidistant manner with increasing dashing distance, the solid line corresponding to $g=0$. (a): $\mr{Re}x$, (b): $\mr{Re}x^d$, (c) $\mr{Im}x$, and (d) $\mr{Im}x^d$ which is vanishing for $g=0$.}\label{depahof}
\end{figure}

\begin{figure}
\includegraphics[scale=0.4]{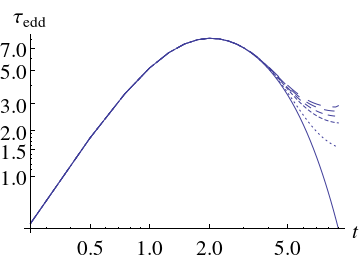}
\caption{The effective dynamical decoherence time, plotted against the time for a slightly overdamped anharmonic oscillator, defined by the boundary conditions $x_i=x_f=x_f^d=0$, $x_i^d=0.5$ and the parameters $\omega_0=0.48$, the coupling constant being distributed in an equidistant manner between $g=0$ and $g=0.024$ with increasing dashing distance and $d_0=d_2=1$.}\label{effdectf}
\end{figure}

Yet another qualitatively new aspect of the anharmonic forces can be found in the numerical quadrature to solve the equations of motion. The simplest possibility is the integration of the equations of motion with a given initial, $\hx(t_i)=\hx_i$, $\dot{\hx}(t_i)=\hat v_i$ or final conditions, $\hx(t_f)=\hx_f$, $\dot{\hx}(t_f)=\hat v_f$, and to adjust the initial or final velocities to satisfy all boundary conditions. Either $x$ or $x^d$ is unstable in these cases which makes the adjustment difficult. Another possibility is to seek each trajectory along its stable time direction but this implies integrating $x$ and $x^d$ in opposite directions in time which leads to new difficulties. The bottom line is that the set of differential equation \eq{eomxxd} together with the boundary condition is stiff for large $t$ and is a challenge to solve numerically. Classical open systems pose no such problem. The typical saddle point, displayed in Fig. \ref{depahof}, shows a significant change as the coupling moves from $g=0$ to $g=0.02$, supporting the enhanced sensitivity of the dynamics for $g\sim0$. The inclusion of an $\ord{x^{d4}}$ anharmonic term in the decoherence potential, $V(x^d)$, tends to decrease this sensitivity. Another message of Fig. \ref{depahof} (b) is that apart of the very weak coupling regime $|\mr{Re}x^d|$ remains bounded by its initial value, $|x_i^d|$. In other words the stationary decoherence strength is an upper bound for the dynamical decoherence. One can introduce an effective dynamical decoherence time scale by retaining the exponential factor in eq. \eq{propdmho},
\be
\frac\hbar{\tau_{edd}}=\frac{\mr{Im}S_{eff}(\hx_f,\hx_i,t)}t,
\ee
which is plotted in Fig. \ref{effdectf}. It shows clearly the singularity at $g=0$ and the slowing down of the double exponential time dependence of the decoherence, corresponding to a straight line on the semi-log plot, to a single exponential due to the non-linear forces. The weakening of the exponential divergence of the dynamical decoherence time scale of the harmonic oscillator to a linear one suggests that the stationary decoherence approximation is applicable, just as in the case of the Brownian motion.

\section{Summary}\label{concls}
The suppression of the interference terms, the decoherence, can be phrased in a dynamical and an instantaneous manner, by considering the interference terms at the initial or at the final time, respectively. The open dynamics is characterized by a local effective Lagrangian. In the case of harmonic dynamics the time dependence is given by the normal frequencies, known from classical physics, the novelty of the quantum level being the opposite orientation of the dissipative force for the coordinate and its quantum fluctuation. The decoherence builds up linearly in time at the beginning of the motion in both schemes with a slope which reflects the initial state and the dynamics. The relaxation in the long time dependence generates a unique characteristic decoherence time scale but the actual suppression is fundamentally different in the two schemes, namely the time dependence is given by a single and a double exponential function for the instantaneous and the dynamical decoherence scheme, respectively. The double exponential function in the decoherence can be traced back to the need of overshooting in constructing the saddle point trajectory for the quantum fluctuations in the presence of a linear restoring force. The dissipation and the decoherence can be separated in harmonic models where the former is realized already at the level of the first moments and the latter appears on that of the second moments only. Anharmonicity changes the picture; it mixes the first two moments, rendering dissipation and decoherence inseparable, ``screens'' the double exponential time dependence of the dynamical decoherence and induces a singularity as $g\to0^+$. 

The Brownian motion, the harmonic and weakly anharmonic oscillators serve as the starting point to approach the physics of more realistic classical dissipative systems. The overshooting of the quantum fluctuation saddle point trajectory of the harmonic oscillator renders the limit $\omega\to0$ of the harmonic oscillator different than the $\omega=0$ Brownian motion. The singularity of the quantum fluctuations at $g=0$ places the harmonic oscillator and the weakly anharmonic oscillator into two, qualitatively different classes of models.

\appendix
\section{Effective Lagrangian}\label{effls}
The effective Lagrangian, \eq{efflagr}, can be derived for a test particle, interacting with an ideal gas environment in the leading order of the system-environment coupling constant \cite{gas}. To this end one assumes a potential, describing the test particle-gas interaction, containing a coupling constant, $g$ and employs the perturbation expansion. The leading order result for the parameters of the Lagrangian are $\ord{g^2}$ expressions, given in terms of a one-loop integral which involves the Lindhard function of the gas. 

An alternative derivation of the effective Lagrangian is based on a harmonic oscillator model, defined by the Lagrangian,
\be
L=\frac{m_B}{2}\dot x^2-U_B(x)+\sum_n\left(\frac{m}2\dot y^2_n-\frac{m\omega_n^2}2y^2_n-g_ny_nx\right)
\ee
with $\omega_n>0$ and $U_B(x)>x^2\sum_ng^2_n/2m_B\omega_n^2$ \cite{calderialeggett}. The influence functional, obtained by the elimination of the environment coordinates, $\hy_n$, can be written in the form $S_{infl}[\hx]=-\hx\hat\Sigma\hx/2$, where $\hat\Sigma=\sum_ng_n^2\hat\sigma\hD_n\hat\sigma/m_B$ denotes the self energy  and $\hat\sigma=\mr{Diag}(1,-1)$  stands for the metric tensor of the simplectic structure, imposed by the time reversal invariance, $S[x^+,x^-]=-S^*[x^-,x^+]$ \cite{grabert,gas}. The propagator of the $n$-th environment coordinate,
\be
\hD_n(t-t')\int\frac{d\omega}{2\pi}e^{-i(t-t')\omega}\hD(\omega,\omega_n),
\ee
is 
\be
\hD(\omega,\Omega)=\begin{pmatrix}\frac1{\omega^2-\Omega^2+i\epsilon}&-i2\pi\Theta(-\omega)\delta(\omega^2-\Omega^2)\cr-i2\pi\Theta(\omega)\delta(\omega^2-\Omega^2)&-\frac1{\omega^2-\Omega^2-i\epsilon}\end{pmatrix}-i\frac{2\pi\delta(\omega^2-\Omega^2)}{e^{\frac{\hbar\Omega}{k_BT}}-1}\begin{pmatrix}1&1\cr1&1\end{pmatrix}
\ee
where the last term describes the the thermal bath effects on the environment attached to. 

The models can conveniently be parameterized by the help of the spectral density,
\be
\rho(\Omega)=\sum_n\frac{g_n^2}{2m_B\omega_n}[\delta(\omega_n-\Omega)-\delta(\omega_n+\Omega)],
\ee
giving the self energy,
\be
\hat\Sigma(\omega)=\frac1{m_B}\int_{-\infty}^\infty d\Omega\Omega\rho(\Omega)\hat\sigma\hD(\omega,\Omega)\hat\sigma.
\ee
The integration over the spectral variable can easily be carried out with the result
\be
\hat\Sigma=\begin{pmatrix}\Sigma^n+i\Sigma^i&\Sigma^f-i\Sigma^i\cr-\Sigma^n-i\Sigma^i&-\Sigma^n+i\Sigma^i\end{pmatrix},
\ee
where 
\be
\Sigma^n(\omega)=2P\int_0^\infty d\Omega\frac{\Omega\rho(\Omega)}{\omega^2-\Omega^2},~~~
\Sigma^f(\omega)=-i\pi\rho(\omega),~~~
\Sigma^i(\omega)=-\pi\rho(|\omega|),
\ee
$P$ denoting the principal value. This is a generic model with arbitrary spectral function. If the environment corresponds to free particles then the space-time symmetries restrict the form of the spectral function up to a multiplicative factor. 

We continue with a oscillator model, defined by the Drude spectral function,
\be
\rho(\Omega)=\frac{\lambda^2}{m_B\Omega_D}\frac{\Omega}{\Omega_D^2+\Omega^2},
\ee
where the self energy is analytical around vanishing frequency. The $\ord{\omega^2}$ contributions define the Lagrangian \eq{efflagr} with a mass and a potential renormalization, $m=m_B+\delta m$, $U(x)=U_B(x)+m_B\delta\omega^2x^2/2$, where $\Delta m=\pi\lambda^2/m_B\Omega_D^4$ and $\delta\omega^2=\pi\lambda^2/m_B^2\Omega_D^2$. The coupling constants to the environment are given in terms of the Drude parameters and the temperature, $k=\pi\lambda^2/m_B\Omega_D^3$, $d_0=2k_BT/\hbar\Omega_D$ and $d_2=\hbar/6k_BT\Omega_D-2k_BT/\hbar\Omega_D^2$.

\acknowledgments
I thank J\'anos Hajdu and Martin Jan\ss en for encouragement and illuminating discussions.

\end{document}